\documentclass[12pt,draftcls,onecolumn]{IEEEtran}

\usepackage{setspace,multirow,tikz,etoolbox}%slashbox
\usetikzlibrary{arrows,automata}
\usepackage{lipsum, enumitem}

\usepackage{blkarray}
\usepackage[T1]{fontenc}
\usepackage[latin9]{inputenc}
\usepackage{amsthm}
\usepackage{amsmath}
\usepackage{graphicx}
\usepackage{amssymb}
\usepackage{url}
\usepackage{verbatim}
\usepackage[export]{adjustbox}
\usepackage{framed,caption}
\usepackage{multirow}

\theoremstyle{plain}
\newtheorem{thm}{Theorem}
\theoremstyle{plain}

\newtoggle{arxiv}
\toggletrue{arxiv}
\newcommand{\arxiv}[1]{\iftoggle{arxiv}{#1}{}}
\newcommand{\noarxiv}[1]{\iftoggle{arxiv}{}{#1}}

\usepackage{cite}\usepackage{amsfonts}\usepackage{times}\usepackage{bm}%\usepackage{dsfont}
\usepackage{amsthm}
\usepackage{subfigure}
\theoremstyle{definition}

\newtheorem{lemma}{Lemma}

\newtheorem*{lemma*}{Lemma}

\theoremstyle{remark}
\newtheorem{remark}{Remark}

\newcommand\uth{u_{\mathrm{th}}}

\title{Optimal Beam Sweeping and Communication in Mobile Millimeter-Wave Networks}

\author{
Nicol\`o Michelusi, Muddassar Hussain
\thanks{N. Michelusi and M. Hussain are with the School of Electrical and Computer Engineering at Purdue University.
Email: \tt{\{michelus,hussai13\}@purdue.edu}.
}
\thanks{This research has been funded by the National Science Foundation under grant CNS-1642982.}
}

\graphicspath{%
     {./Figures/}
}

\begin{document}
\maketitle

\begin{abstract}
Millimeter-wave (mm-wave) communications incur a high beam alignment cost in mobile scenarios such as vehicular networks. Therefore, an efficient beam alignment mechanism is required to mitigate the resulting overhead. In this paper, a one-dimensional mobility model is proposed where a mobile user (MU), such as a vehicle, moves along a straight road with time-varying and random speed, and communicates with base stations (BSs) located on the roadside over the mm-wave band. To compensate for location uncertainty, the BS widens its transmission beam and, when a critical beamwidth is achieved, it performs beam-sweeping to refine the MU position estimate, 
followed by data communication over a narrow beam. The average rate and average transmission power are computed in closed form and the optimal beamwidth for communication, number of sweeping beams, and transmission power allocation are derived so as to maximize the average rate under an average power constraint. Structural properties of the optimal design are proved, and a bisection algorithm to determine the optimal sweeping -- communication parameters is designed. It is shown numerically that an adaptation of the IEEE 802.11ad standard to the proposed model exhibits up to 90\% degradation in spectral efficiency compared to the proposed scheme. 
\end{abstract}
%\vspace{-5mm}
\section{Introduction}
%Vehicular sensing capabilities such as GPS, radar, cameras, Lidar \cite{levinson}, have witnessed tremendous improvements over the last few decades \cite{lu}. Sharing of the data from vehicle sensors between different vehicles and roadside units (RSU) provides enhancements in sensing range as well as the redundancy in case of sensors failure. Moreover, vehicular communication can enable a wide range of infotainment services such as digital maps, cloud computing, ultra high definition video streaming, etc. To enable these services, data rates on the order of Gbps are required, which cannot be supported by current cellular or dedicated short range communication (DSRC) standards \cite{kenney,araniti}. 

Millimeter-wave (mm-wave) technology has emerged as a promising solution to enable multi-Gbps communication, thanks to the abundant bandwidth available \cite{choi}. 
Mm-wave will be key to supporting autonomous transportation systems by 
allowing vehicles to extend their sensing range and make more informed decisions
by exchanging rich sensory information \cite{va}. It will also enable a wide range of infotainment services such as digital maps, cloud computing, ultra-high definition video streaming, etc.
However, signal propagation at these frequencies poses several challenges to the design of future communication systems supporting high throughput and high mobility, such as high isotropic path loss and sensitivity to blockages \cite{rappaport_book}.
Mm-wave systems are expected to leverage narrow-beam communications to counteract the propagation loss \cite{channel_model} by using large antenna arrays at both base stations (BSs) and mobile users (MUs).

However, sharp beams are susceptible to beam mis-alignment due to mobility or blockage, 
necessitating frequent re-alignment. This task
can be challenging, especially in mobile scenarios.
The beam alignment protocol may consume time, frequency, and energy resources, thus potentially offsetting the benefits of mm-wave directionality. Therefore, it is imperative to 
design schemes to mitigate its overhead.

In this paper, we investigate the trade-off between beam alignment and data communication in mobile mm-wave networks. We propose a beam-sweeping -- data communication protocol that accounts for the uncertainty on the location and speed of the MU and for the temporal overhead of beam-sweeping.
Specifically, the BS associated with the MU
widens its transmission beam to compensate for the increasing uncertainty on the MU location and, when a critical beamwidth is achieved,
it performs beam-sweeping to refine the MU's position estimate and achieve a narrow communication beam.
We compute the performance in closed-form, and investigate the design of the optimal
beamwidth for communication, number of sweeping beams, and transmission power
so as to maximize the average rate under average power constraint.
We find structural properties and propose a bisection method to determine the optimal design. We show numerically that an adaptation of IEEE 802.11ad to our model
exhibits a performance degradation up to $90\%$ compared to our design.

\par Beam alignment in mm-wave has been a subject of intensive research due to its importance in mm-wave  communication systems. The research in this area can be categorized into beam-sweeping \cite{exhaustive,iterative,bisection,asilomar17}; AoA/AoD estimation \cite{alkhateeb,marzi}; and data-assisted schemes \cite{radar,lowfreq,va,inverse_finger}. 
Beam-sweeping based schemes require scanning of regions of uncertainty of AoA/AoD. The simplest and yet most popular form of beam-sweeping is the so-called \emph{exhaustive} search \cite{exhaustive}, which sequentially scans through all possible beam pairs from the BS and MU codebooks and selects the one with maximum signal power. This approach has been adopted in existing mm-wave standards including IEEE 802.15.3c  \cite{ieee80215c} and IEEE 802.11ad \cite{ieee80211ad}. The other popular scheme is a hierarchical form of scanning called \emph{iterative} search \cite{iterative}, where beam-sweeping is first performed using wider beams followed by refinement with narrow beams. In our previous work \cite{asilomar17}, we derived an energy-efficient scheme termed \emph{fractional search}, which minimizes the energy consumption subject to a rate  constraint: in each slot, the BS adaptively scans a fraction of the uncertainty region of the AoD, function of the slot index, rate requirement, probabilities of false-alarm and mis-detection, bandwidth, path loss, and noise power spectral density. 

\par AoA/AoD estimation aims to reduce the number of measurements required by beam-sweeping by leveraging the sparsity of mm-wave channels, \emph{e.g.}, via compressive sensing as in \cite{alkhateeb}.  The paper \cite{marzi} derived an approximate maximum likelihood estimator for the channel by directly exploiting the structure of  the mm-wave channel.
Data-aided schemes utilize information from 
radar \cite{radar}, lower frequencies \cite{lowfreq}, or positional information \cite{inverse_finger,va}
to reduce the cost of beam-sweeping.
 Based on this idea, the authors of \cite{va} proposed a beamwidth optimization algorithm that maximizes the data rate for non-overlapping beams. In contrast to \cite{va}, we propose an analytical framework for the joint optimization of beamwidth, communication power and beam-sweeping to maximize the communication performance.
\emph{To the best of our knowledge, we are the first to propose an analytical framework for the optimization of the beam-sweeping and communication parameters in mobile mm-wave networks.}

\begin{figure}
    \centering% \scriptsize
    \includegraphics[width=.9\linewidth,trim = 0mm 0mm 0mm 0mm,clip=false]{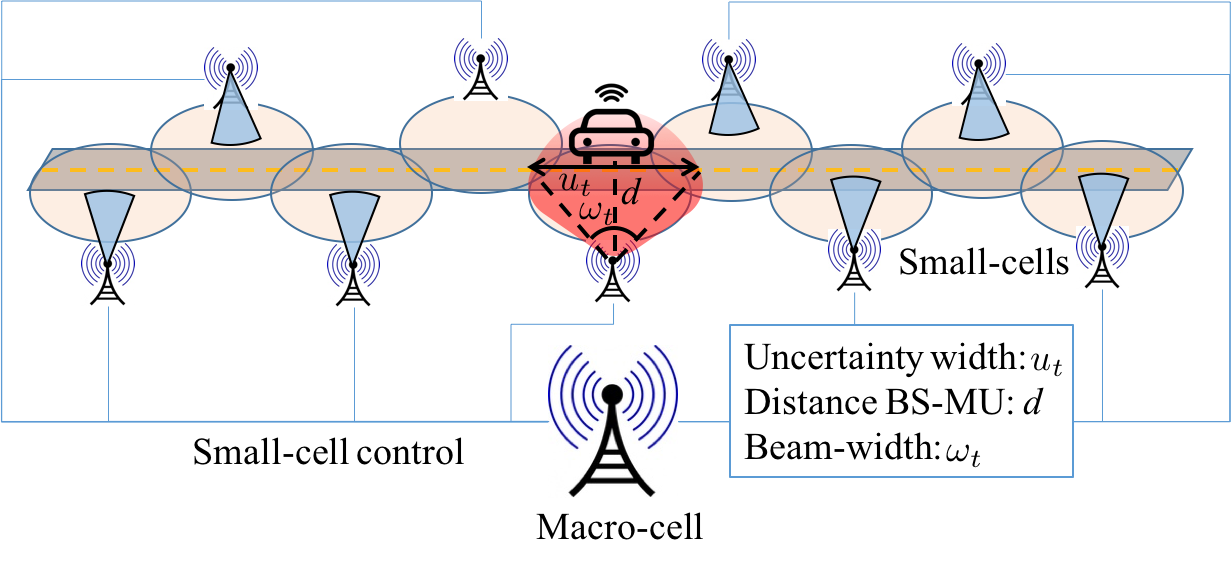}
    %\vspace{-2mm}
\caption{System model.}
\label{figexlabel}
%\vspace{-6mm}
\end{figure}

The paper is organized as follows: in Sec. \ref{sysmo}, we present the system model
and optimization problem;
in Sec. \ref{analysis}, we present the analysis, followed by numerical results in Sec. \ref{numres};  finally, in Sec. \ref{conclu}, we conclude with some remarks.
\noarxiv{Due to space constraints, the analytical proofs are provided in \cite{proofshere}.}

%\vspace{-2mm}
\section{System Model}
\label{sysmo}
We consider a \emph{dense} cell deployment, as depicted in Fig.~\ref{figexlabel}. The MU is associated with its closest BS, at distance $d$.
We assume that the BS points its beam perpendicularly to the motion of the MU (a good approximation in dense cell deployments).
A macro-cell unit controls functions such as handover among cells. The time-scale of this task is larger than the beam-sweeping -- data communication cycle, and thus we neglect it.
We neglect the additional overhead due to channel estimation, Doppler correction, and
the impact of  beamwidth on Doppler spread  (see \cite{7742901}).
%\vspace{-5mm}
\subsection{User mobility model}
The MU moves along a line (\emph{e.g.}, a vehicle along a road).  Let $(p_t,v_t)\in\mathbb R^2$ be its position and speed at time $t$.
We assume that $v_t\in[v_{\min},v_{\max}]$, where $v_{\min}<v_{\max}$ (possibly, negative), and we let $v_{\mathrm{drift}}=(v_{\min}+v_{\max})/2$ be the drift velocity and $\phi\triangleq v_{\max}-v_{\min}$ be the speed uncertainty.
$v_t$ is time-varying and random, with \emph{arbitrary} distribution in $[v_{\min},v_{\max}]$.
The speed parameters $v_{\mathrm{drift}},\phi$ are assumed to be known,
and can be estimated from GPS information collected at the macro-cell
(\emph{e.g.}, via lower frequency dedicated short range communication channels \cite{kenney}).
Herein, we assume that $v_{\mathrm{drift}}=0$, since a known non-zero drift can be incorporated by appropriate beam steering.
Thus, it follows that $v_t\in[-\phi/2,\phi/2]$ and, given $p_0$ at a reference time $0$, 
\begin{align}
\label{mobmodel}
p_t=p_0+\int_0^t v_{\tau}\mathrm d\tau\in \left[p_0-\frac{\phi t}{2},p_0+\frac{\phi t}{2}\right].
\end{align}

In this paper, the uncertainty on the location of the MU at time $t$ is denoted by the \emph{uncertainty interval}
$\mathcal U_t{\equiv}[\hat p_{t}{-}u_{t}/2,\hat p_{t}{+}u_{t}/2]$, where $\hat p_t$ is the median estimated position and 
$u_t$ is the \emph{uncertainty width}, so that $p_t\in\mathcal U_t$.
From the mobility model (\ref{mobmodel}), if no beam-sweeping is done in the time interval $[t,\tau]$, the uncertainty width augments at rate $\phi$, \emph{i.e.},
\begin{align}
\label{widthmodel}
u_\tau=u_{t}+\phi(\tau-t),\ \tau\geq t,
\end{align}
and is reduced via beam-sweeping, as discussed in Sec. \ref{beamsweep}.

The communication between BS and MU follows a beam-sweeping -- data communication cycle of duration $T$. We now describe the entire cycle, starting from the reference time $t{=}0$.
%\vspace{-5mm}
\subsection{Beam Sweeping}
\label{beamsweep}
When, at the reference time $t=0$, the uncertainty width reaches a critical value $u_0=\uth$,
the BS currently associated with the MU
 sweeps the entire uncertainty interval $\mathcal U_0$ using $\eta\geq 2,\eta\in\mathbb N$  beams,
 transmitted sequentially over $\eta$ microslots, each of duration $\delta_S$.
 During this interval, the uncertainty width increases over time due to MU mobility. In order to compensate for it,
 the BS scans wider regions over successive microslots, as detailed below.
 Thus, we let $\omega_{i}$ be the beamwidth of the $i$th beam, where $i=1,2,\dots,\eta$.
 
 At the end of the beam-sweeping interval of duration $\eta\delta_S$, the MU processes the signals, and feeds back  to the BS the ID of the strongest signal (\emph{e.g.}, via a lower frequency control channel).  
The BS uses such strongest beam to communicate with the MU in the data communication phase,
as detailed in Sec. \ref{datacomm}.
We neglect the time to send this feedback signal.

$\{\omega_i,i=1,2,\dots,\eta\}$ are designed with the following requirements:
{\bf R1} -- By the end of the beam-sweeping phase,
the entire uncertainty interval $\mathcal U_0$ must be scanned,
plus the additional uncertainty resulting from the MU mobility during the beam-sweeping phase; {\bf R2} -- the beamwidth at the beginning of the data communication phase, $u_{\eta\delta_S}$, must be independent
of the strongest beam selected.

To guarantee {\bf R2}, note  that,
  if the $i$th beam, $i=1,2,\dots,\eta$ is the strongest one detected (with beamwidth $\omega_{i}$),
the uncertainty width at the end of the beam-sweeping phase becomes\footnote{Herein, we assume that $\omega_i\ll 2\pi$, so that 
the length of the interval scanned in the $i$th microslot is
$2d\tan(\omega_i/2)\simeq d\omega_i$, see Fig. \ref{figexlabel}
(the beam is approximated as being pointed perpendicularly to the motion of the MU).} 
\begin{align}
u_{\eta\delta_S}=d\omega_{i}+(\eta+1-i)\delta_S\phi,
\end{align}
due to the MU mobility in the subsequent $(\eta+1-i)$ microslots until the end of beam-sweeping. Hence, {\bf R2} requires
\begin{align}
\label{omegait}
\omega_{i}=\omega_{1}+(i-1)\frac{\delta_S\phi}{d},\ \forall i=1,2,\dots,\eta,
\end{align}
so that, at the end of beam-sweeping, the uncertainty width becomes
\begin{align}
\label{attheend}
u_{\eta\delta_S}=d\omega_{1}+\eta\delta_S\phi,\ \forall i.
\end{align}

We now discuss how to design $\omega_1$ (and $\omega_i$ via (\ref{omegait})) so as to guarantee {\bf R1}.
At the reference time $0$, the uncertainty interval is $[0,\uth]$.
 In the first microslot, the BS scans the interval $[0,d\omega_{1}]$ using a beam with beamwidth $\omega_{1}$.
 If the MU is within this interval, at the end of the 
 beam-sweeping phase it will detect the ID of the strongest beam as $\#1$,
and the uncertainty width will thus be given by (\ref{attheend}).
 Otherwise (if the MU is outside of this interval),  after the first microslot the MU may be in the interval 
 $[d\omega_{1}-\delta_S\phi/2,\uth+\delta_S\phi/2]$, which accounts for the additional uncertainty
 due to the MU mobility in the time interval $[0,\delta_S]$.
 Thus, in the second microslot, the BS scans the interval
 $[d\omega_{1}-\delta_S\phi/2,d\omega_{1}+d\omega_{2}-\delta_S\phi/2]$
using a beam with beamwidth  $\omega_{2}$.
   If the MU is within this interval, at the end of the beam sweeping phase it will detect the ID of the strongest beam as $\#2$, and the uncertainty width will thus be given by (\ref{attheend}).
 Otherwise (if the MU is outside of this interval),  after the second microslot the MU may be in the interval 
  $[d\omega_{1}+d\omega_{2}-\delta_S\phi,\uth+\delta_S\phi]$, which accounts for the additional uncertainty
 due to the MU mobility in the time interval $[\delta_S,2\delta_S]$.
  Thus, in the third microslot, the BS scans the interval
  $[d\omega_{1}+d\omega_{2}-\delta_S\phi,d\omega_{1}+d\omega_{2}+d\omega_{3}-\delta_S\phi]$
  with a beam with beamwidth equal to  $\omega_{3}$, and so on.
  
  By induction, at the beginning of the $i$th microslot, where $i=1,2,...,\eta$, $i-1$ beams have been scanned. 
  If the MU was located within one of the previous $i-1$ beams (say the $j$th, $j\leq i-1$), it will detect the ID of the strongest beam as $\#j$ at the end of the beam-sweeping phase, and the uncertainty width will thus be given by (\ref{attheend}).
 Otherwise (if the MU is located within one of the next beams $i,i+1,\dots,\eta$), the MU may be in the interval 
 $[d\sum_{k=1}^{i-1}\omega_{k}-(i-1)\delta_S\phi/2,\uth+(i-1)\delta_S\phi/2]$
 at the beginning of the $i$th microslot, which accounts for the additional uncertainty
 due to the MU mobility in the time interval $[0,(i-1)\delta_S]$.
   Thus, in the $i$th microslot, the BS scans the interval
   $[d\sum_{k=1}^{i-1}\omega_{k}-(i-1)\delta_S\phi/2,d\sum_{k=1}^{i}\omega_{k}-(i-1)\delta_S\phi/2]$
  using a beam with beamwidth $\omega_{i}$.
       If the MU is within this interval, it will detect the ID of the strongest beam as $\#i$
 at the end of the beam-sweeping period, and the uncertainty width will thus be given by (\ref{attheend}).
 Otherwise (if the MU is outside of this interval),  at the end of the $i$th microslot the MU may be in the interval 
  $[d\sum_{k=1}^{i}\omega_{k}-i\delta_S\phi/2,\uth+i\delta_S\phi/2]$, which accounts for the additional uncertainty
 due to the MU mobility in the time interval $[(i-1)\delta_{S},i\delta_S]$.
 
Using a similar argument, in the last microslot (the $\eta$th one), if the MU was not located within one of the
 previous $\eta-1$ beams, then the MU will be located in the interval
 $[d\sum_{k=1}^{\eta-1}\omega_{k}-(\eta-1)\delta_S\phi/2,\uth+(\eta-1)\delta_S\phi/2]$
 of width $\uth+(\eta-1)\delta_S\phi-d\sum_{k=1}^{\eta-1}\omega_{k}$.
 This must be scanned exhaustively with a beam of width $\omega_{\eta}$, 
 hence
 \begin{align}
 \label{omegaetat}
d\omega_{\eta}=\uth+(\eta-1)\delta_S\phi-d\sum_{k=1}^{\eta-1}\omega_{k}.
 \end{align}
 By combining (\ref{omegaetat}) with (\ref{omegait}) we obtain,
 $\forall i=1,2,\dots,\eta,$
 \begin{align}
\omega_{i}
=\frac{\uth}{d\eta}- \frac{(\eta-1)(\eta-2)}{2\eta}\frac{\delta_S\phi}{d}+(i-1)\frac{\delta_S\phi}{d}.
\end{align}
  At the end of beam-sweeping, data communication begins
  and  the new uncertainty width is given by (\ref{attheend}), yielding
  \begin{align}
  u_{\mathrm{comm}}(\uth,\eta)
{\triangleq}u_{\eta\delta_S}{=}
\frac{\uth}{\eta}{+}\eta\delta_S\phi{-}\delta_S\phi\frac{(\eta{-}1)(\eta{-}2)}{2\eta}.
\end{align}
which evolves over the data communication interval according to (\ref{widthmodel}).

Note that a feasible beam is such that  $\omega_{k}\geq 0,\forall k=1,2,\dots,\eta$.
Additionally, beam-sweeping must reduce the uncertainty width, \emph{i.e.}, $u_{\mathrm{comm}}(\uth,\eta)\leq \uth$. These two conditions together yield
\begin{align}
\label{constuth}
\uth\geq
\delta_S\phi\max\left\{\frac{\eta^2/2+3/2\eta-1}{\eta-1},\frac{1}{2}(\eta-1)(\eta-2)\right\}.
\end{align}
Herein, we assume that the correct sector is detected with no error by the MU (this requires proper beam design to achieve small false-alarm and misdetection probabilities, see \cite{mmnets17}). 
\subsection{Data communication}
\label{datacomm}
Immediately after beam-sweeping, at time $t=\eta\delta_S$, the data communication phase begins,
 and the uncertainty width is $u_{\eta\delta_S}=u_{\mathrm{comm}}(\uth,\eta)$.
 The uncertainty width $u_t$ increases over time due to the mobility of the MU,
 according to (\ref{widthmodel}). The data communication period,
 and the beam-sweeping -- data communication cycle, terminate 
 at time $T$ such that $u_{T}=\uth$, at which time a new cycle begins.
 From (\ref{widthmodel}) we obtain 
\begin{align}
T=\frac{(\eta-1)\uth}{\phi\eta}+\frac{\delta_S}{2}\frac{(\eta-1)(\eta-2)}{\eta}.
%%%
\end{align}
In the time interval $[\eta\delta_S,T]$,
the transmission beam of the BS associated with the MU
is chosen so as to support reliable communication over the entire uncertainty interval.
Its beamwidth is thus chosen as
$\omega_t\simeq u_t/d$ [rad].\footnote{Note that we assume that $u_t/d\ll 2\pi$, so that we can approximate the beamwidth as $\omega_t=2\arctan(u_t/d/2)\simeq u_t/d$, see Fig. \ref{figexlabel}.} 
\begin{remark}
Note that in our model the beamwidth is varied continuously within a continuous set 
$\omega\in[u_{\mathrm{comm}}(\uth,\eta)/d,\uth/d]$ for analytical tractability.
This approach is a continuous approximation of a practical deployment where
 the system may operate at discrete times 
 using a discrete codebook to generate transmission beams with different beamwidths \cite{noh}.
\end{remark}
Let $P_t$ be the transmission power per Hz at time $t$ to communicate reliably.
Assuming isotropic reception at the MU \cite{bisection,asilomar17}, the instantaneous 
transmission rate is given by
\begin{align}
\label{rate}
&R_t=W_{\mathrm{tot}}\log_2\left(1+\gamma\frac{P_t}{\omega_t}\right),
\end{align}
where $W_{\mathrm{tot}}$ is the bandwidth,
 $\gamma\triangleq \frac{\lambda^2\xi}{8\pi d^2 N_0 W_{\mathrm{tot}}}$ is the SNR scaling factor,
 $\lambda$ is the wavelength, $N_0$ is the noise power spectral density,
 and $\xi$ is the antenna efficiency.
 Note that $P_t$ is spread evenly across the angular directions covered by the transmission beams,
so that $P_t/\omega_t$ is the power per radian delivered to the receiver.

%\vspace{-2mm}
\subsection{Performance metrics and optimization problem}
The optimal choice of the beam-sweeping and communication parameters reflects a trade-off between locating the MU with high accuracy so as to achieve narrow-beam communication, and mitigating the overhead in terms of sweeping time. This is the goal of our design.

Let $\eta\geq 2,\eta\in\mathbb N$, $\uth$ satisfying (\ref{constuth}), and $P:[\eta\delta_S,T]\mapsto \mathbb R_+$ be the transmit power function in the data communication phase.
We define the time-average communication rate and transmission power, defined over one
beam-sweeping -- data communication cycle $[0,T]$, as
\begin{align}
&\!\!\!\!\bar R(\eta,\uth,P){=}\frac{W_{\mathrm{tot}}}{T}\!\!\int_{\eta\delta_S}^{T}\!\!\!\!\log_2\!\!\left(1{+}\frac{d\gamma P_t}{
u_{\mathrm{comm}}(\uth,\eta){+}\phi t}\right)\!\mathrm d t,\!\!
\\
&\!\!\!\!\bar P(\eta,\uth,P){=}\frac{W_{\mathrm{tot}}}{T}\int_{\eta\delta_S}^{T}P_t\mathrm d t.
\end{align}
The goal is to determine the optimal design of 
the joint data communication and beam-sweeping parameters 
$(\eta,\uth,P)$ so as to maximize the average rate under average power constraint $P_{\max}>0$,
\emph{i.e.},
\begin{align}
{\bf P1:}\quad(\eta,\uth,P)^*=&\underset{(\eta,\uth,P)}{\arg\max}
\ \bar R(\eta,\uth,P),
\\&
\text{s.t. }\bar P(\eta,\uth,P)\leq P_{\max}.
\end{align}
The analysis is carried out in the next section.
\section{Analysis}
\label{analysis}
Due to the concavity of the $\log_2$ function, Jensen's inequality yields the following result.
\begin{lemma}
The optimal power allocation function $P:[\eta\delta_S,T]\mapsto \mathbb R_+$ is given by the water-filling scheme
\begin{align}
\label{Pu}
&P_t=\left(\rho-\frac{u_t}{d\gamma}\right)^+,\ \forall t\in[\eta\delta_S,T],
\end{align}
where $\rho\geq \frac{u_{\mathrm{comm}}(\uth,\eta)}{d\gamma}$ is a parameter to optimize.\qed
\end{lemma}
Under the water-filling power allocation, the design space is simplified to 
$(\eta,\uth,\rho)$, where $\eta\geq 2,\eta\in\mathbb N$, $\uth$ satisfies (\ref{constuth}),
and $\rho\geq\frac{u_{\mathrm{comm}}(\uth,\eta)}{d\gamma}$.
The average rate and average transmission power can be computed in closed form and are given by\footnote{We replace the dependence on the power allocation function $P$ with the parameter $\rho$.}
\begin{align}
\nonumber
\label{avgR2}
&\bar R(\eta,\uth,\rho)=
   \frac{W_{\mathrm{tot}}}{\ln(2)\phi T}
   \Bigg[
   \Big(\uth{-}u_{\mathrm{comm}}(\uth,\eta)\Big)\Big(1{+}\ln\left(d\gamma\rho\right)\Big)
\\&
\nonumber
- \uth\ln(\uth)
+u_{\mathrm{comm}}(\uth,\eta)\ln(u_{\mathrm{comm}}(\uth,\eta))
\\&
+\chi(d\gamma\rho\leq\uth)\left(\uth\ln\left(\frac{\uth}{d\gamma\rho}\right)+d\gamma\rho-\uth\right)
\Bigg]
,
%%%%
% OK CHECKED!!!!
%%%%
\\&
\label{avgP2}
\bar P(\eta,\uth,\rho)
=
\chi(d\gamma\rho\leq\uth)\frac{(\uth-d\gamma\rho)^2}{2d\phi\gamma T}
\\&
+\frac{\uth-u_{\mathrm{comm}}(\uth,\eta)}{2d\phi\gamma T}
\Big(2d\gamma\rho-\uth-u_{\mathrm{comm}}(\uth,\eta)\Big),
\nonumber
%%%%
% OK CHECKED
%%%%
\end{align}
where $\chi(\cdot)$ denotes the indicator function.

It is useful to define the following change of variables:
\begin{align}
&\upsilon\triangleq\frac{\uth}{\delta_S\phi},
\\
&\zeta\triangleq\frac{d\gamma\rho}{\delta_S\phi\upsilon}-1
\geq
\frac{u_{\mathrm{comm}}(\uth,\eta)}{\delta_S\phi\upsilon}-1.
\end{align}
The performance metrics (\ref{avgR2})-(\ref{avgP2}) can thus be expressed as
\begin{align}
&\hat u_{\mathrm{comm}}(\upsilon,\eta)
\triangleq
\frac{u_{\mathrm{comm}}(\uth,\eta)}{\delta_S\phi}
=
\frac{\upsilon}{\eta}+\frac{1}{2}\eta+\frac{3}{2}-\frac{1}{\eta},
\\
\nonumber
\label{avgR}
&
\hat R(\eta,\upsilon,\zeta)
\triangleq
\frac{\ln(2)}{W_{\mathrm{tot}}}\bar R(\eta,\uth,\rho)
=
\frac{\eta}{\eta-1}\frac{1}{\upsilon+\frac{\eta}{2}-1}
\\&
\times\Bigg[
\Big(\upsilon-\hat u_{\mathrm{comm}}(\upsilon,\eta)\Big)\Big(1{+}\ln(1+\zeta)\Big)
\\&
{-}\hat u_{\mathrm{comm}}(\upsilon,\eta)
\ln\left(\frac{\upsilon}{\hat u_{\mathrm{comm}}(\upsilon,\eta)}\right)
\nonumber
{+}\chi(\zeta{<}0)\upsilon\left(\zeta{-}\ln(1{+}\zeta)\right)
\!\!\Bigg],
%%%%
% OK CHECKED!!!!
%%%%
\\&
\label{avgP}
\hat P(\eta,\upsilon,\zeta)
{\triangleq}
\frac{d\gamma}{\delta_S\phi}\bar P(\eta,\uth,\rho)
{=}
\frac{\eta\upsilon^2\zeta^2\chi(\zeta{<}0)}{2(\eta-1)\big(\upsilon+\frac{\eta}{2}-1\big)}
\nonumber
\\&
{+}\frac{\eta(\upsilon-\hat u_{\mathrm{comm}}(\upsilon,\eta))}{2(\eta-1)\big(\upsilon+\frac{\eta}{2}-1\big)}
\Big(2\upsilon(1+\zeta){-}\upsilon{-}\hat u_{\mathrm{comm}}(\upsilon,\eta)\Big),
%%%%
% OK CHECKED
%%%%
\end{align}
where  (\ref{constuth}) and $\rho\geq \frac{u_{\mathrm{comm}}(\uth,\eta)}{d\gamma}$ yield
the feasible set
\begin{align}
\nonumber
\mathcal F_{\eta}\equiv&
\Big\{
(\upsilon,\zeta):
\upsilon\geq\upsilon_{\min}(\eta),
\zeta\geq\frac{\hat u_{\mathrm{comm}}(\upsilon,\eta)}{\upsilon}-1
\Big\},
\end{align}
and we have defined 
\begin{align}
\upsilon_{\min}(\eta)\triangleq\max\left\{\frac{\eta^2+3\eta-2}{2(\eta-1)},\frac{1}{2}(\eta-1)(\eta-2)\right\}.
\end{align}
Note that we have normalized the average rate and transmission power,
so that they no longer depend on the system parameters $W_{\mathrm{tot}},\phi,d,\gamma,\delta_S$.
This is beneficial since it unifies the structure of the optimal design in a wide range of scenarios.

%\nm{There exists $\eta^*>4$ such that  $\frac{\frac{1}{2}[\eta^2+3\eta-2]+\hat\delta_F\eta}{\eta-1}>\frac{1}{2}(\eta-1)(\eta-2)$ for $\eta<\eta^*$ and contrary otherwise ($\eta^*=4.5$ if $\hat\delta_F=0$).}

The optimization problem thus becomes
\begin{align}
{\bf P2:}\quad(\eta,\upsilon,\zeta)^*=&\underset{\eta\geq 2,\eta\in\mathbb N,(\upsilon,\zeta)\in\mathcal F_\eta}{\arg\max}
\ \hat R(\eta,\upsilon,\zeta)
\\&
\text{s.t. }\hat P(\eta,\upsilon,\zeta)\leq \hat P_{\max},
\end{align}
where $\hat P_{\max}=\frac{d\gamma}{\delta_S\phi}P_{\max}$.
This optimization problem is non-convex. 
We have the following structural result.
\begin{thm}
\label{thm1}
$\zeta<0$ is suboptimal.\noarxiv{\qed}
\end{thm}
\arxiv{
\begin{proof}
See Appendix \ref{proofofthm1}.
\end{proof}}
The intuition behind Theorem \ref{thm1} is that, if $\zeta<0$, then the water-filling power allocation
is such that $P_t=0$ during a portion of the data communication phase.
 This is suboptimal: it is more energy-efficient to reduce 
the beam-sweeping threshold $\uth$ and increase $\zeta$ so as to reduce the "idle" time interval in the communication phase.

Thus, in the following we focus on the case  $\zeta\geq0$. 
Note that $\hat P(\eta,\upsilon,\zeta)$ needs to satisfy the power constraint. Since
it is an increasing function of $\zeta$, 
we must have $\hat P(\eta,\upsilon,0)\leq \hat P_{\max}$ to obtain a feasible solution, yielding
\begin{align}
\label{dgn}
\!\upsilon{\leq}\frac{\eta^2{+}3\eta{-}2}{2(\eta-1)}{+}\frac{\eta\hat P_{\max}}{\eta-1}
\left(1+\sqrt{1+\frac{2\eta}{\hat P_{\max}}}\right)\triangleq \upsilon_{\max}(\eta).\!\!
% OK CHECKED
%%%%
\end{align}
Note that $\upsilon$ must also satisfy the constraint
$\upsilon\geq\upsilon_{\min}(\eta)$,
hence we must have $\upsilon_{\max}(\eta)\geq\upsilon_{\min}(\eta)$.
If $\eta\leq 4$, then $\frac{1}{2}\frac{\eta^2+3\eta-2}{\eta-1}>\frac{1}{2}(\eta-1)(\eta-2)$
and any $\upsilon $ satisfying (\ref{dgn}) also satisfies $\upsilon\geq\upsilon_{\min}(\eta)$. 
On the other hand, if $\eta\geq 5$ then $\frac{1}{2}\frac{\eta^2+3\eta-2}{\eta-1}<\frac{1}{2}(\eta-1)(\eta-2)$
and $\upsilon_{\max}(\eta)\geq\upsilon_{\min}(\eta)$ is equivalent to
\begin{align}
\hat P_{\max}\geq\frac{1}{2}\frac{[\eta^2-5\eta+2]^2}{\eta^2-4\eta+2},\text{ for }\eta\geq 5.
\end{align}
Since the right hand side is an increasing function of $\eta\geq 5$, 
we conclude that
there exists $4\leq\eta_{\max}<\infty$
such that the problem is feasible for all $2\leq\eta\leq\eta_{\max}$ (indeed, the problem is always feasible for $\eta\in\{2,3,4\}$ since $\upsilon_{\max}(\eta)\geq\upsilon_{\min}(\eta)$ in this case).
We thus define the new feasibility set  as
\begin{align*}
\mathcal F\equiv
\left\{(\upsilon,\eta):
2\leq\eta\leq\eta_{\max},\eta\in\mathbb N,
\upsilon_{\min}(\eta)\leq \upsilon\leq \upsilon_{\max}(\eta)
\right\}.
\end{align*}
Let $(\upsilon,\eta)\in\mathcal F$.
Under such pair, $\hat P(\eta,\upsilon,\zeta)$ and $\hat R(\eta,\upsilon,\zeta)$ are  increasing functions of $\zeta\geq 0$, hence the optimal $\zeta$ is such that
 the power constraint 
is attained with equality. We thus obtain $\zeta$ as a function of $(\upsilon,\eta)$ as
\begin{align}
\zeta(\upsilon,\eta)
\triangleq
\frac{(\eta-1)(\upsilon+\eta/2-1)}{\eta\upsilon[\upsilon-\hat u_{\mathrm{comm}}(\upsilon,\eta)]}
(\hat P_{\max}-\hat P(0,\eta,\upsilon)).
\end{align}

Since the power constraint is satisfied with equality for 
$(\upsilon,\eta)\in\mathcal F$ and $\zeta=\zeta(\upsilon,\eta)$,
 the optimization problem becomes unconstrained, yielding
 \begin{align}
{\bf P3:}\quad(\eta,\upsilon)^*=&\underset{(\upsilon,\eta)\in\mathcal F}{\arg\max}
\ \hat R(\eta,\upsilon,\zeta(\upsilon,\eta)),
\end{align}
and $\zeta^*=\zeta(\upsilon^*,\eta^*)$,
where 
\begin{align}
&\hat R(\eta,\upsilon,\zeta(\upsilon,\eta))
=
\frac{\eta}{\eta-1}\frac{1}{\upsilon+\frac{\eta}{2}-1}
\\&
\times\Bigg[
\Big(\upsilon-\hat u_{\mathrm{comm}}(\upsilon,\eta)\Big)\Big(1{+}\ln(1+\zeta(\upsilon,\eta))\Big)
\\&
- \hat u_{\mathrm{comm}}(\upsilon,\eta)
\ln\left(\frac{\upsilon}{\hat u_{\mathrm{comm}}(\upsilon,\eta)}\right)
\nonumber
\Bigg].
\end{align}

We solve the optimization problem as follows: for each $2\leq\eta\leq\eta_{\max}$,
we solve 
\begin{align}
\label{maxRgiveneta}
\upsilon^*(\eta)
=\underset{\upsilon_{\min}(\eta)\leq\upsilon\leq\upsilon_{\max}(\eta)}{\arg\max}
\hat R(\eta,\upsilon,\zeta(\upsilon,\eta)).
\end{align}
Then, the optimal $\eta^*$ and $\upsilon^*$ are found by optimizing $\eta$ via exhaustive search over the finite discrete set $\{2,3,\dots,\eta_{\max}\}$,
\begin{align}
\eta^*=\arg\max_{\eta\in\{2,3,\dots,\eta_{\max}\}}
\hat R(\rho(\upsilon^*(\eta),\eta),\eta,\upsilon^*(\eta)),
\end{align}
and $\upsilon^*=\upsilon^*(\eta^*)$.

\subsection{Solution of (\ref{maxRgiveneta}) given $\eta\in\{2,3,\dots,\eta_{\max}\}$}
In this section, we investigate how to compute $\upsilon^*(\eta)$ given $\eta\in\{2,3,\dots,\eta_{\max}\}$.
We have the following theorem.
\begin{thm}
\label{thm2}
Given $\eta\in\{2,3,\dots,\eta_{\max}\}$, the optimal $\upsilon^*(\eta)$ is given by
\begin{align}
\label{optupseta}
\upsilon^*(\eta)=\max\left\{\frac{1}{2}(\eta-1)(\eta-2),\hat\upsilon\right\},
\end{align}
where $\hat\upsilon$ is the unique solution in $(\frac{1}{2}\frac{\eta^2+3\eta-2}{\eta-1},\upsilon_{\max}(\eta))$ of 
$f_{\eta}(\upsilon)=0$, where
\begin{align}
\nonumber
\label{fups}
&f_\eta(\upsilon)
\triangleq
- \frac{\upsilon-\hat u_{\mathrm{comm}}(\upsilon,\eta)}{\upsilon(1+\rho(\upsilon,\eta))}\frac{(\eta-1)(\upsilon+\eta/2-1)+2\eta}{2\eta}
\\&
- \frac{(\eta-1)(\upsilon+\eta/2-1)}{\eta(1+\rho(\upsilon,\eta))}\rho(\upsilon,\eta)
\\&
+\ln(1+\rho(\upsilon,\eta))\eta+\ln(\upsilon/\hat u_{\mathrm{comm}}(\upsilon,\eta))(\eta/2+1).\noarxiv{\qed}
\nonumber
\end{align}
\end{thm}
\arxiv{
\begin{proof}
See Appendix \ref{proofofthm2}.
\end{proof}}

The function $f_\eta(\upsilon)$ is proportional to the derivative of $\hat R(\eta,\upsilon,\zeta(\upsilon,\eta))$ with respect to $\upsilon$, up to a positive multiplicative factor.
Note that $\hat\upsilon$ can be determined using the bisection method. In fact,
$f_\eta(\upsilon)$ is a decreasing function of $\upsilon$ (see proof of the theorem in \cite{proofshere}),
with 
\begin{align}
\label{limits}
\lim_{\upsilon\to\frac{1}{2}\frac{\eta^2+3\eta-2}{\eta-1}}f_\eta(\upsilon)=\infty
\text{ and }
f_\eta(\upsilon_{\max}(\eta))<0.
\end{align}

\begin{table}[b]
\footnotesize
\centering
\caption{Simulation parameters}
\label{tb}
\begin{tabular}{|r|c|l|}
\hline
 Parameter&Symbol &Value  \\ \hline
 Carrier frequency & $f_c$& 60 GHz\\
 Bandwidth &$W_{\mathrm{tot}}$ & 1.76 GHz\\
 Noise PSD & $N_0$ &  -174 dBm/Hz\\
 Microslot duration &$\delta_S$ & 10$\mu$s\\
 Distance BU-MU & $d$ & 10 m\\
 Antenna efficiency & $\xi$ & 1\\
 \hline
\end{tabular}
\end{table}

\section{Numerical Results}
\label{numres}
In this section, we present numerical results to demonstrate the performance of the proposed beam-sweeping -- data communication protocol. We compare our proposed scheme with an adaptation of IEEE 802.11ad to our model, in which partially overlapping beams of $7^o$ beamwidth are employed such that adjacent beams share half of the beam area. Moreover, to evaluate this scheme we assume a worst-case scenario where the vehicle moves with either speed of $v_{\max}$ or $v_{\min}=-v_{\max}$. Therefore, with IEEE 802.11ad, beam alignment is required after each $r/v_{\max}$ [s] (the time required for the MU to move to the edge of the beam), where $r = d \tan\left(\frac{7^o}{2} \right)$. Once the edge of the beam is reached (thus, the MU is located in either position $p\in\{-r,r\}$),
the BS scans the two beams covering the intervals $[-2r,0]$ and $[0,2r]$, each with 
$7^o$ beamwidth,
so that the time overhead of beam sweeping is $2\delta_S$. Immediately after,
the strongest beam is detected and data communication proceeds.
Then, the fraction of time spent in data communication is given as
\begin{align}
f_{\mathrm{comm}} = \frac{r/v_{\max}}{r/v_{\max}+2\delta_S},
\end{align}
 the average throughput of IEEE 802.11ad is given as
\begin{align}
\bar R_{11\mathrm{ad}} =W_{\mathrm{tot}}\log_2 \left( 1 + \gamma\frac{P_t}{7\pi/180} \right) \times f_{\mathrm{comm}},
\end{align}
and the average power as  $\bar P_{11\mathrm{ad}}=P_t\times f_{\mathrm{comm}}$. 
The common parameters of the simulation are given in Table \ref{tb}. 

\begin{figure}[!t]
\centering
\includegraphics[width=.9\columnwidth]{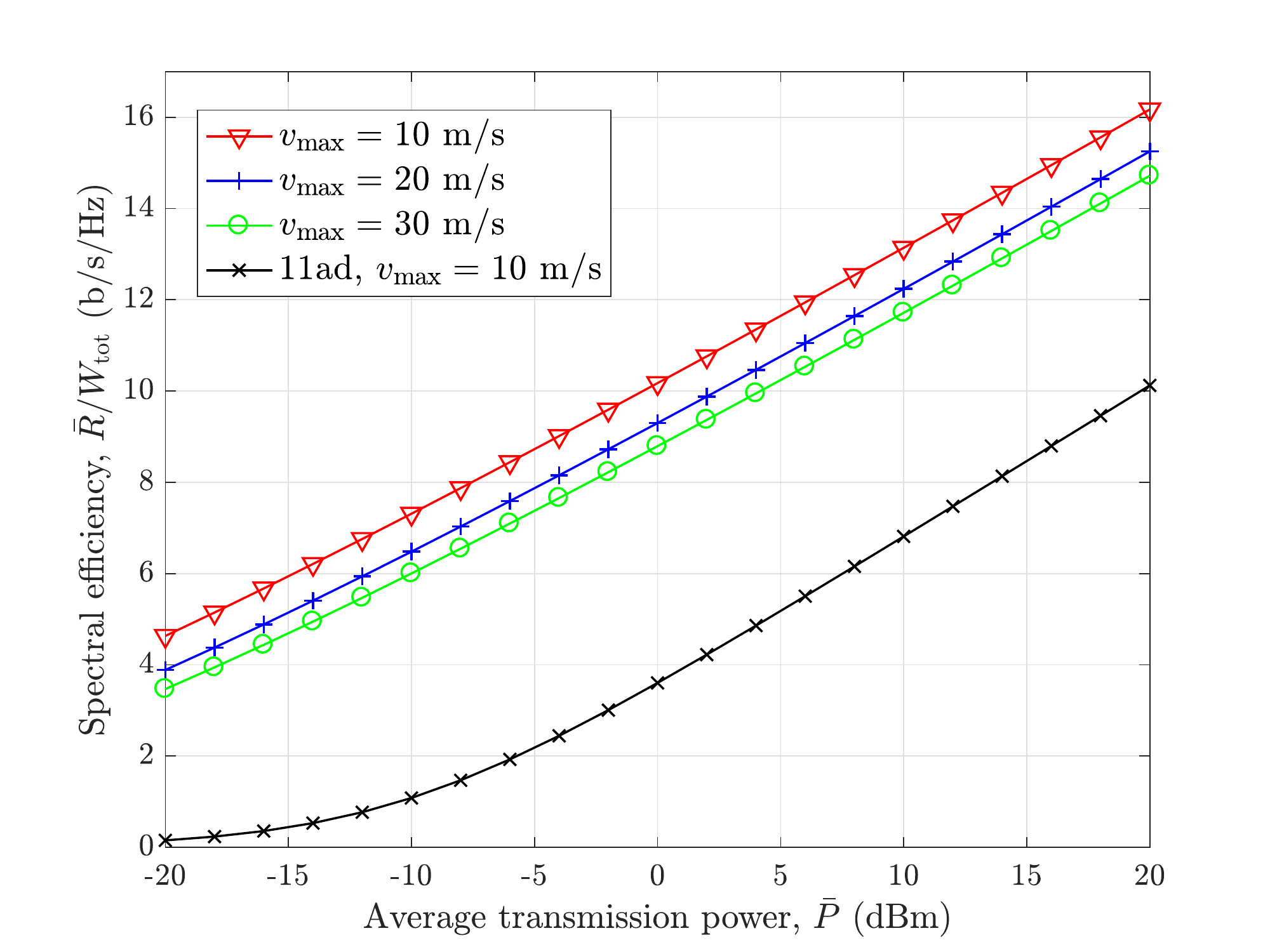}
\caption{Average spectral efficiency versus average power.}
\label{fig:rate_power}
%\vspace{-5mm}
\end{figure}

\par In Fig. \ref{fig:rate_power}, we plot the average spectral efficiency $\bar R/W_{\mathrm{tot}}$ versus the average power consumption $\bar P$. A monotonic trend between the spectral efficiency and the average power is observed. Moreover, the performance of the system deteriorates as we increase the speed, due to the increasing overhead of beam alignment. Additionally, we observe that IEEE 802.11ad performs poorly since it uses fixed $7^o$ beams which are not optimized to the specific mobile scenario, with degradation up to $90\%$  compared to our proposed scheme. 

\par
\begin{figure}[!t]
\centering
\includegraphics[width=.9\columnwidth]{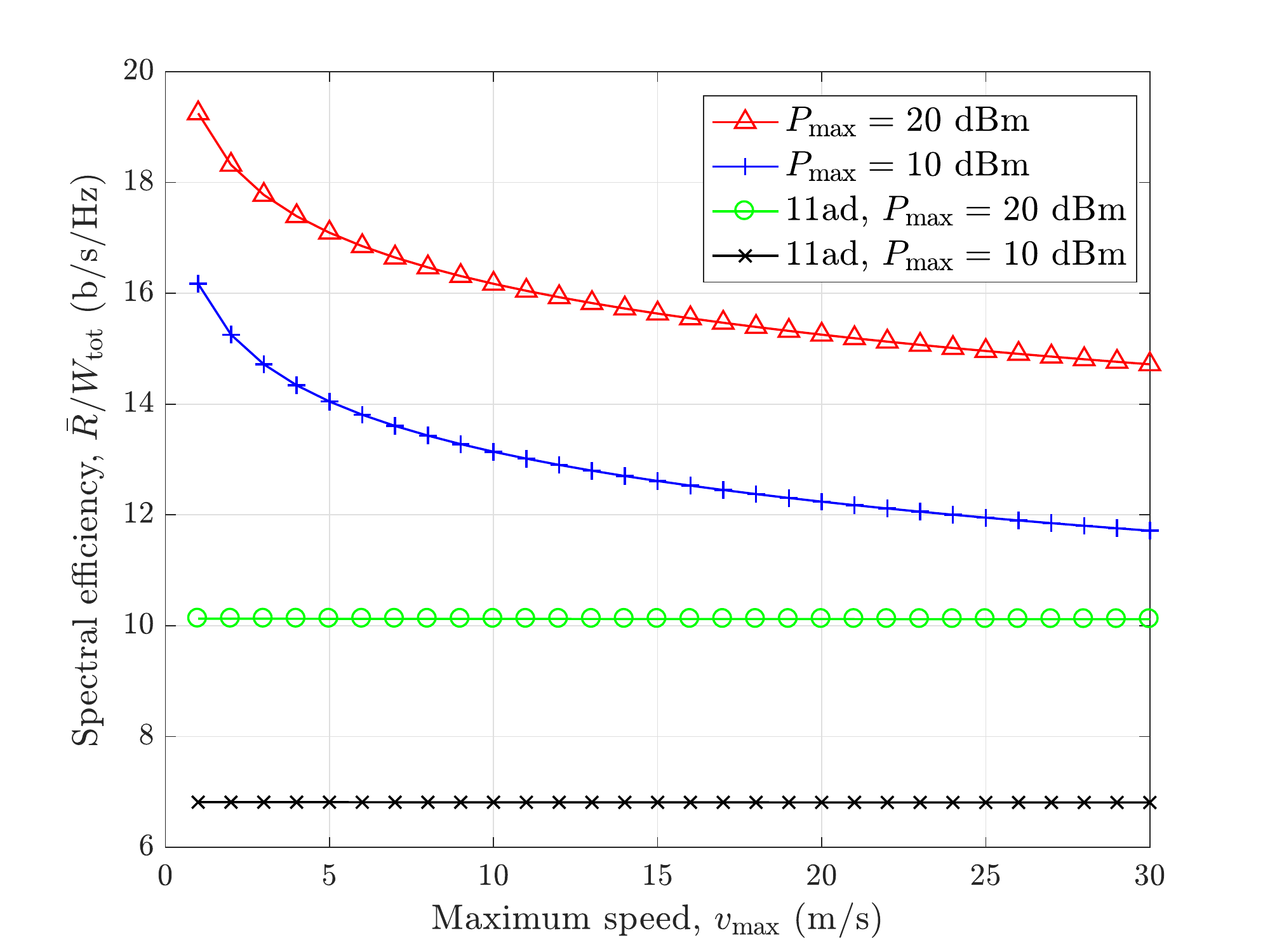}
\caption{Average spectral efficiency versus speed.}
\label{fig:rate_speed}
%\vspace{-5mm}
\end{figure}

 In Fig.~\ref{fig:rate_speed}, we plot the effect of speed on the spectral efficiency for two different values of the average power $\bar P$. It can be seen that the 
spectral efficiency of the proposed scheme degrades monotonically as the speed $v_{\max}$ is increased. Moreover, the performance improves with higher value of $\bar P$ as observed also in Fig. \ref{fig:rate_power}. It can be noticed that the curves corresponding to IEEE 802.11ad do not show significant degradation as the speed is increased.
This is due to the relatively wide beam used in IEEE 802.11ad, so that beam alignment is relatively infrequent.
 However, the performance of IEEE 802.11ad is poor compared to  our proposed scheme.
 %\vspace{-5mm}
\section{Conclusion}
\label{conclu}
In this paper, we propose a one-dimensional mobility model where a vehicle moves along a straight road with time-varying and random speed and communicates with base stations located on the roadside over the mm-wave band. 
We propose a beam-sweeping -- data communication protocol and study its performance in closed form.
We derive structural properties of the optimal design, based on which we design a bisection algorithm.
We compare numerically our proposed design to an adaptation 
of IEEE 802.11ad to our model, which exhibits performance degradation up to $90\%$.
%\vspace{-5mm}

\bibliographystyle{IEEEtran}
\bibliography{IEEEabrv,biblio} 

\begin{thebibliography}{10}
\providecommand{\url}[1]{#1}
\csname url@samestyle\endcsname
\providecommand{\newblock}{\relax}
\providecommand{\bibinfo}[2]{#2}
\providecommand{\BIBentrySTDinterwordspacing}{\spaceskip=0pt\relax}
\providecommand{\BIBentryALTinterwordstretchfactor}{4}
\providecommand{\BIBentryALTinterwordspacing}{\spaceskip=\fontdimen2\font plus
\BIBentryALTinterwordstretchfactor\fontdimen3\font minus
  \fontdimen4\font\relax}
\providecommand{\BIBforeignlanguage}[2]{{%
\expandafter\ifx\csname l@#1\endcsname\relax
\typeout{** WARNING: IEEEtran.bst: No hyphenation pattern has been}%
\typeout{** loaded for the language `#1'. Using the pattern for}%
\typeout{** the default language instead.}%
\else
\language=\csname l@#1\endcsname
\fi
#2}}
\providecommand{\BIBdecl}{\relax}
\BIBdecl

\bibitem{choi}
J.~Choi, V.~Va, N.~Gonzalez-Prelcic, R.~Daniels, C.~R. Bhat, and R.~W. Heath,
  ``Millimeter-wave vehicular communication to support massive automotive
  sensing,'' \emph{IEEE Communications Magazine}, vol.~54, no.~12, pp.
  160--167, December 2016.

\bibitem{va}
V.~Va, T.~Shimizu, G.~Bansal, and R.~W. Heath, ``Beam design for beam switching
  based millimeter wave vehicle-to-infrastructure communications,'' in
  \emph{2016 IEEE International Conference on Communications (ICC)}, May 2016,
  pp. 1--6.

\bibitem{rappaport_book}
T.~S. Rappaport, \emph{{Wireless communications: principles and
  practice}}.\hskip 1em plus 0.5em minus 0.4em\relax Prentice Hall PTR, 2002.

\bibitem{channel_model}
M.~R. Akdeniz, Y.~Liu, M.~K. Samimi, S.~Sun, S.~Rangan, T.~S. Rappaport, and
  E.~Erkip, ``{Millimeter Wave Channel Modeling and Cellular Capacity
  Evaluation},'' \emph{IEEE Journal on Selected Areas in Communications},
  vol.~32, no.~6, pp. 1164--1179, June 2014.

\bibitem{exhaustive}
C.~Jeong, J.~Park, and H.~Yu, ``{Random access in millimeter-wave beamforming
  cellular networks: issues and approaches},'' \emph{IEEE Communications
  Magazine}, vol.~53, no.~1, pp. 180--185, January 2015.

\bibitem{iterative}
V.~Desai, L.~Krzymien, P.~Sartori, W.~Xiao, A.~Soong, and A.~Alkhateeb,
  ``{Initial beamforming for mmWave communications},'' in \emph{48th Asilomar
  Conference on Signals, Systems and Computers}, Nov 2014, pp. 1926--1930.

\bibitem{bisection}
M.~Hussain and N.~Michelusi, ``{Throughput optimal beam alignment in millimeter
  wave networks},'' in \emph{Information Theory and Applications Workshop
  (ITA)}, Feb 2017, pp. 1--6.

\bibitem{asilomar17}
------, ``Energy efficient beam alignment in millimeter wave networks,'' in
  \emph{2017 Asilomar Conference on Signals, Systems, and Computers}, 2017, to
  appear.

\bibitem{alkhateeb}
A.~Alkhateeb, O.~E. Ayach, G.~Leus, and R.~W. Heath, ``Channel estimation and
  hybrid precoding for millimeter wave cellular systems,'' \emph{IEEE Journal
  of Selected Topics in Signal Processing}, vol.~8, no.~5, pp. 831--846, Oct
  2014.

\bibitem{marzi}
Z.~Marzi, D.~Ramasamy, and U.~Madhow, ``Compressive channel estimation and
  tracking for large arrays in mm-wave picocells,'' \emph{IEEE Journal of
  Selected Topics in Signal Processing}, vol.~10, no.~3, pp. 514--527, April
  2016.

\bibitem{radar}
N.~González-Prelcic, R.~Méndez-Rial, and R.~W. Heath, ``Radar aided beam
  alignment in mmwave v2i communications supporting antenna diversity,'' in
  \emph{Information Theory and Applications Workshop (ITA)}, Jan 2016, pp.
  1--7.

\bibitem{lowfreq}
T.~Nitsche, A.~B. Flores, E.~W. Knightly, and J.~Widmer, ``Steering with eyes
  closed: Mm-wave beam steering without in-band measurement,'' in \emph{2015
  IEEE Conference on Computer Communications (INFOCOM)}, April 2015, pp.
  2416--2424.

\bibitem{inverse_finger}
V.~Va, J.~Choi, T.~Shimizu, G.~Bansal, and R.~W. Heath, ``{Inverse Multipath
  Fingerprinting for Millimeter Wave V2I Beam Alignment},'' \emph{IEEE
  Transactions on Vehicular Technology}, vol.~PP, no.~99, pp. 1--1, 2017.

\bibitem{ieee80215c}
``{IEEE} {Std} 802.15.3c-2009,'' \emph{IEEE Standard}, pp. 1--200, Oct 2009.

\bibitem{ieee80211ad}
``{IEEE} {Std} 802.11ad-2012,'' \emph{IEEE Standard}, pp. 1--628, Dec 2012.

\bibitem{7742901}
V.~Va, J.~Choi, and R.~W. Heath, ``{The Impact of Beamwidth on Temporal Channel
  Variation in Vehicular Channels and Its Implications},'' \emph{IEEE
  Transactions on Vehicular Technology}, vol.~66, no.~6, pp. 5014--5029, June
  2017.

\bibitem{kenney}
J.~B. Kenney, ``Dedicated short-range communications (dsrc) standards in the
  united states,'' \emph{Proceedings of the IEEE}, vol.~99, no.~7, pp.
  1162--1182, July 2011.

\bibitem{mmnets17}
M.~Hussain, D.~J. Love, and N.~Michelusi, ``{Neyman-Pearson Codebook Design for
  Beam Alignment in Millimeter-Wave Networks},'' in \emph{Proceedings of the
  1st ACM Workshop on Millimeter-Wave Networks and Sensing Systems}.\hskip 1em
  plus 0.5em minus 0.4em\relax New York, NY, USA: ACM, 2017, pp. 17--22.

\bibitem{noh}
S.~Noh, M.~D. Zoltowski, and D.~J. Love, ``{Multi-Resolution Codebook and
  Adaptive Beamforming Sequence Design for Millimeter Wave Beam Alignment},''
  \emph{IEEE Transactions on Wireless Communications}, vol.~16, no.~9, pp.
  5689--5701, Sept 2017.

\bibitem{proofshere}
N.~Michelusi and M.~Hussain, ``{Optimal Beam Sweeping and Communication in
  Mobile Millimeter-wave Networks},'' Purdue University, Tech. Rep., 2017,
  \url{https://engineering.purdue.edu/~michelus/ICC2018.pdf}.

\end{thebibliography}

\arxiv{
\appendices
\section{Proof of Theorem \ref{thm1}}
\label{proofofthm1}
\begin{proof}
First, note that if $\zeta=\hat u_{\mathrm{comm}}(\upsilon,\eta)/\upsilon-1$, then
$\hat P(\upsilon(1+\zeta),\eta,\upsilon)=0<P_{\max}$ and $\hat R(\upsilon(1+\zeta),\eta,\upsilon)=0$. This configuration is clearly suboptimal since a non-zero rate
can be achieved by increasing $\zeta$.

Now, let $\hat u_{\mathrm{comm}}(\upsilon,\eta)/\upsilon-1<\zeta<0$ and assume this configuration is optimal.
Note that this implies  $\upsilon>\hat u_{\mathrm{comm}}(\upsilon,\eta)$, or equivalently
$\upsilon>\frac{1}{2}\frac{\eta^2+3\eta-2}{\eta-1}$.

We have two cases: 1) $\upsilon>\frac{1}{2}(\eta-1)(\eta-2)$ and 2) $\upsilon=\frac{1}{2}(\eta-1)(\eta-2)$ (and consequently $\eta\geq 5$
since we must also have $\upsilon>\frac{1}{2}\frac{\eta^2+3\eta-2}{\eta-1}$).

\paragraph{$\upsilon>\frac{1}{2}(\eta-1)(\eta-2)$}
We show that, by 
increasing $\zeta$ and decreasing $\upsilon$ so as to preserve the power consumption, the rate strictly increases, 
and thus we achieve a contradiction. From (\ref{avgR}) and (\ref{avgP}) with $\zeta<0$ we obtain
\begin{align}
\label{Rtot3}
&
\hat R(\eta,\upsilon,\zeta)
=
\frac{\eta}{\eta-1}\frac{\hat u_{\mathrm{comm}}(\upsilon,\eta)}{\upsilon+\frac{\eta}{2}-1}
\\&
\times\Bigg[
\frac{\upsilon}{\hat u_{\mathrm{comm}}(\upsilon,\eta)}(1+\zeta)-1
- \ln\left(\frac{\upsilon}{\hat u_{\mathrm{comm}}(\upsilon,\eta)}(1+\zeta)\right)
\nonumber
\Bigg],
%%%%
% OK CHECKED!!!!
%%%%
\\&
\label{Ptot3}
\hat P(\eta,\upsilon,\zeta)
=
\frac{\eta
\Big(\upsilon(1+\zeta)-\hat u_{\mathrm{comm}}(\upsilon,\eta)\Big)^2
}{2(\eta-1)\big(\upsilon+\frac{\eta}{2}-1\big)}.
\end{align}
We increase $\zeta$ by $h>0$ (arbitrarily small) and decrease $\upsilon$ by a function $g(h)>0$, so as to maintain the power consumption unaltered, \emph{i.e.},
\begin{align}
\hat P(\eta,\upsilon,\zeta)=\hat P(\eta,\upsilon-g(h),\zeta+h).
\end{align}
In the limit $h\to0$ we must have
\begin{align}
\label{x2}
\frac{\mathrm d\hat P(\eta,\upsilon,\zeta)}{d\zeta}
-g^\prime(0)\frac{\mathrm d\hat P(\eta,\upsilon,\zeta)}{d\upsilon}=0,
\end{align}
where $g^\prime(0)$ is the derivative of $g(h)$ in zero, which must be positive since $g(h)>0$ for arbitrarily small $h$.
To show this, note that
\begin{align}
&\frac{\mathrm d\hat P(\zeta,\eta,\upsilon)}{d\zeta}
=
\frac{\eta\upsilon\Big(\upsilon(1+\zeta)-\hat u_{\mathrm{comm}}(\upsilon,\eta)\Big)}{(\eta-1)\big(\upsilon+\frac{\eta}{2}-1\big)}>0,
\\
\nonumber
&\frac{\mathrm d\hat P(\zeta,\eta,\upsilon)}{d\upsilon}
=
\frac{\eta\Big(\upsilon(1+\zeta)-\hat u_{\mathrm{comm}}(\upsilon,\eta)\Big)
}{2(\eta-1)\big(\upsilon+\frac{\eta}{2}-1\big)^2}
\\&
\times
\Big(
(1+\zeta)(\upsilon+\eta-2)
- \frac{\upsilon}{\eta}+\frac{1}{2}\eta+\frac{1}{2}+\frac{1}{\eta}
\Big)
>0,
%%%%
\end{align}
where the last inequality follows from the fact that
$\zeta>\hat u_{\mathrm{comm}}(\upsilon,\eta)/\upsilon-1$.
Hence, it follows that indeed $g^\prime(0)>0$.

We now show that, for arbitrarily small $h$,
\begin{align}
\hat R(\eta,\upsilon,\zeta)<\hat R(\eta,\upsilon-g(h),\zeta+h).
\end{align}
Equivalently, in the limit $h\to0$, we must have
\begin{align}
\label{x1}
\frac{\mathrm d\hat R(\eta,\upsilon,\zeta)}{d\zeta}
-g^\prime(0)\frac{\mathrm d\hat R(\eta,\upsilon,\zeta)}{d\upsilon}
>0.
\end{align}
Note that 
\begin{align}
\label{x}
&\frac{\mathrm d\hat R(\eta,\upsilon,\zeta)}{d\zeta}
=
\frac{1}{\upsilon(1+\zeta)}\frac{\mathrm d\hat P(\zeta,\eta,\upsilon)}{d\zeta},
\\
% DONE
&\frac{\mathrm d\hat R(\zeta,\eta,\upsilon)}{d\upsilon}
=
\frac{\eta(\eta/2-1)}{(\eta-1)\upsilon(\upsilon+\frac{\eta}{2}-1)^2}
[\upsilon(1+\zeta)-\hat u_{\mathrm{comm}}(\upsilon,\eta)]
\nonumber
\\&
+\frac{\eta(\eta/2+1)}{(\eta-1)(\upsilon+\frac{\eta}{2}-1)^2}
\ln\left(\frac{\upsilon}{\hat u_{\mathrm{comm}}(\upsilon,\eta)}(1+\zeta)\right),
% OK CHECKED!
\end{align}
and thus replacing (\ref{x}) in (\ref{x1}), and using (\ref{x2}) and the fact that $g^\prime(0)>0$, we obtain the equivalent condition
\begin{align}
\frac{\mathrm d\hat P(\eta,\upsilon,\zeta)}{d\upsilon}
-\upsilon(1+\zeta)\frac{\mathrm d\hat R(\eta,\upsilon,\zeta)}{d\upsilon}
>0,
\end{align}
iff
\begin{align}
\label{x3}
\nonumber
&g(\zeta)\triangleq\Big(1-\frac{\hat u_{\mathrm{comm}}(\upsilon,\eta)}{\upsilon(1+\zeta)}\Big)
\Big(
(1+\zeta)\upsilon- \frac{\upsilon}{\eta}+\frac{1}{2}\eta+\frac{1}{2}+\frac{1}{\eta}
\Big)
\\&
-(\eta+2)\ln\left(\frac{\upsilon}{\hat u_{\mathrm{comm}}(\upsilon,\eta)}(1+\zeta)\right)
>0,
\end{align}
which we are now going to prove.
The derivative with respect to $\zeta$ is given by
\begin{align}
\nonumber
&\frac{\mathrm dg(\zeta)}{\mathrm d\zeta}
\propto
\eta(\upsilon(1+\zeta)-\hat u_{\mathrm{comm}}(\upsilon,\eta))^2
\\&
+(2\upsilon+\eta-2)(\upsilon(1+\zeta)-\hat u_{\mathrm{comm}}(\upsilon,\eta))
>0,
\end{align}
where the inequality follows from the fact that 
$\zeta>\hat u_{\mathrm{comm}}(\upsilon,\eta)/\upsilon-1$.
It follows that $g(\zeta)$ is an increasing function of $\zeta$, minimized at $\zeta=\hat u_{\mathrm{comm}}(\upsilon,\eta)/\upsilon-1$, thus proving the inequality.

\paragraph{$\upsilon=\frac{1}{2}(\eta-1)(\eta-2)$ and $\eta\geq 5$}
In this case, we cannot decrease $\upsilon$ any further.
Using a similar approach as in the previous case, we show that a strictly larger rate can be obtained by decreasing both $\eta$ and $\zeta$,
while preserving the power consumption.
From (\ref{Rtot3}) and (\ref{Ptot3}) with $\upsilon=\frac{1}{2}(\eta-1)(\eta-2)$, we obtain
\begin{align}
%\upsilon=\eta^2/2-3/2\eta+1
%&\hat u_{\mathrm{comm}}(\upsilon,\eta)=\eta
\label{Rtot4}
&\hat R(\eta,\upsilon,\zeta)
=
1+\zeta
\\&
 -\frac{2\eta}{(\eta-1)(\eta-2)}\left[1+\ln\left(\frac{(\eta-1)(\eta-2)(1+\zeta)}{2\eta}\right)\right],
 \nonumber
%%%%
% OK CHECKED!!!!
%%%%
\\&
\label{Ptot4}
\hat P(\eta,\upsilon,\zeta)
=
\frac{\left[\upsilon(1+\zeta)-\hat u_{\mathrm{comm}}(\upsilon,\eta)\right]^2}{(\eta-1)(\eta-2)},
%%%%
% OK CHECKED
%%%%
\end{align}
where $\frac{-\eta^2+5\eta-2}{(\eta-1)(\eta-2)}<\zeta<0$.

Now, we decrease $\eta$ by one unit, while keeping $\upsilon$ as before,
and we choose the new $\zeta$, denoted as $\hat \zeta$,
in such a way as to preserve the power consumption. Note that
$\upsilon\geq\upsilon_{\min}(\eta-1)$ hence the constraint on $\upsilon$ is still satisfied,
since $\upsilon_{\min}(\eta-1)$ is a decreasing function of $\eta$.

From (\ref{Ptot3}) with $\upsilon=\frac{1}{2}(\eta-1)(\eta-2)$ we obtain
\begin{align}
%%%%
% OK CHECKED!!!!
%%%%
\label{Ptot5}
\hat P(\eta-1,\upsilon,\hat\zeta)
=
\frac{(\eta-1)
\Big(\upsilon(1+\hat\zeta)-\hat u_{\mathrm{comm}}(\upsilon,\eta-1)\Big)^2
}{(\eta-2)(\eta^2-2\eta-1)}.
%%%%
% OK CHECKED
%%%%
\end{align}
where
\begin{align}
\label{dgfdg}
\hat u_{\mathrm{comm}}(\upsilon,\eta-1)=
\eta-\frac{1}{\eta-1}
<
\hat u_{\mathrm{comm}}(\upsilon,\eta)=\eta.
\end{align}
$\hat\zeta$ is chosen so that $\hat P(\eta-1,\upsilon,\hat\zeta)
=\hat P(\eta,\upsilon,\zeta)$, yielding
\begin{align}
\nonumber
&\upsilon(1+\hat\zeta)
=
\hat u_{\mathrm{comm}}(\upsilon,\eta-1)
\\&
+\sqrt{\eta^2-2\eta-1}
\frac{\upsilon(1+\zeta)-\hat u_{\mathrm{comm}}(\upsilon,\eta)}{\eta-1}.
\end{align}
Thus, it follows that $\upsilon(1+\hat\zeta)>\hat u_{\mathrm{comm}}(\upsilon,\eta-1)$
Additionally, using  (\ref{dgfdg}) 
and the fact that $\upsilon(1+\zeta)-\hat u_{\mathrm{comm}}(\upsilon,\eta)>0$
it follows that
 $\upsilon(1+\hat\zeta)<\upsilon(1+\zeta)$. Therefore
\begin{align}
\hat u_{\mathrm{comm}}(\upsilon,\eta-1)<\upsilon(1+\hat\zeta)<\upsilon(1+\zeta)<\upsilon,
\end{align}
since $\zeta<0$, hence $
\hat u_{\mathrm{comm}}(\upsilon,\eta-1)/\upsilon-1<\hat\zeta<0$.
We now show that this new configuration strictly increases the throughput. From (\ref{Rtot3})
and using the expression of $\hat\zeta$ and of $\hat u_{\mathrm{comm}}(\upsilon,\eta-1)$ we obtain
\begin{align}
\label{Rtot5}
&\hat R(\eta-1,\upsilon,\hat\zeta)
=
\frac{2}{(\eta-2)\sqrt{\eta^2-2\eta-1}}(\upsilon(1+\zeta)-\eta)
\\&
\nonumber
-\frac{2(\eta^2-\eta-1)}{(\eta-2)(\eta^2-2\eta-1)}\ln(\upsilon(\hat\zeta+1)/\hat u_{\mathrm{comm}}(\upsilon,\eta-1)),
\end{align}
and therefore
\begin{align}
\label{Rtot6}
&h(\zeta)\triangleq\hat R(\eta-1,\upsilon,\hat\zeta)-\hat R(\eta,\upsilon,\zeta)
\\&=\nonumber
 \frac{4(\upsilon(1+\zeta)-\eta)}{(\eta-1)(\eta-2)\sqrt{\eta^2-2\eta-1}(\eta-1+\sqrt{\eta^2-2\eta-1})}
 \\& \nonumber
+\frac{2\eta}{(\eta-1)(\eta-2)}\ln(\upsilon(1+\zeta)/\eta)
 \\& \nonumber
-\frac{2(\eta^2-\eta-1)}{(\eta-2)(\eta^2-2\eta-1)}\ln(\upsilon(\hat\zeta+1)/\hat u_{\mathrm{comm}}(\upsilon,\eta-1)).
\end{align}
The derivative of $h(\zeta)$ with respect to $\zeta$ is given by
\begin{align}
\nonumber
&\frac{\mathrm dh(\zeta)}{\mathrm d\zeta}
\propto
\upsilon(1+\zeta)-\hat u_{\mathrm{comm}}(\upsilon,\eta)
\\&
+\frac{2(\upsilon(1+\zeta)-\eta)^2}{\eta-1+\sqrt{\eta^2-2\eta-1}}>0.
%%%%
\end{align}
Therefore, $h(\zeta)$ is an increasing function of $\zeta$, minimized at 
$\zeta=\hat u_{\mathrm{comm}}(\upsilon,\eta)/\upsilon-1$,
yielding
\begin{align}
&\hat R(\hat\zeta,\eta-1,\upsilon)-\hat R(\zeta,\eta,\upsilon)
>
0.
\end{align}
The Theorem is thus proved.
\end{proof}

\section{Proof of Theorem \ref{thm2}}
\label{proofofthm2}
\begin{proof}
To study the optimization problem {\bf P3}, we study the derivative 
$\hat R(\eta,\upsilon,\zeta(\upsilon,\eta))$ with respect to $\upsilon$. We have that
\begin{align}
\nonumber
&\frac{\mathrm d\hat R(\eta,\upsilon,\zeta(\upsilon,\eta))}{\mathrm d\upsilon}
=
\frac{\mathrm d\hat R(\eta,\upsilon,\zeta)}{\mathrm d\upsilon}
+\frac{\mathrm d\hat R(\eta,\upsilon,\zeta)}{\mathrm d\zeta}
\frac{\mathrm d\zeta(\upsilon,\eta)}{\mathrm d\upsilon}
\\&
\propto f_\eta(\upsilon),
\end{align}
where $\propto$ denotes proportionality up to a positive multiplicative factor,
with $f_\eta(\upsilon)$ given by (\ref{fups}).
Therefore, $\hat R(\zeta,\eta,\upsilon)$ is an increasing function 
of $\upsilon$ iff $f_\eta(\upsilon)>0$. We now show that 
$f_\eta(\upsilon)$ is a strictly decreasing function of $\upsilon$,
with limits given by (\ref{limits}).

Note that $\hat\upsilon$ can be determined using the bisection method. In fact,
$f_\eta(\upsilon)$ is a decreasing function of $\upsilon$
 (see proof of the theorem),
with 
\begin{align}
\label{limups}
\lim_{\upsilon\to\frac{1}{2}\frac{\eta^2+3\eta-2}{\eta-1}}f_\eta(\upsilon)=\infty
\end{align}
and 
\begin{align}
f_\eta(\upsilon_{\max}(\eta))<0
\end{align}
(see second part of the proof). Therefore, there exists a unique $\hat\upsilon\in(\frac{1}{2}\frac{\eta^2+3\eta-2}{\eta-1},\upsilon_{\max}(\eta))$ such that 
$f_\eta(\hat\upsilon)=0$, and $\frac{\mathrm d\hat R(\eta,\upsilon,\zeta(\upsilon,\eta))}{\mathrm d\upsilon}>0$ for $\upsilon<\hat\upsilon$
and 
$\frac{\mathrm d\hat R(\eta,\upsilon,\zeta(\upsilon,\eta))}{\mathrm d\upsilon}<0$ for $\upsilon>\hat\upsilon$.

Therefore, the maximum of $\hat R(\eta,\upsilon,\zeta(\upsilon,\eta))$ with respect to
$\upsilon\in(\frac{1}{2}\frac{\eta^2+3\eta-2}{\eta-1},\upsilon_{\max}(\eta))$ is attained at 
$\upsilon=\hat\upsilon$. By combining this result with the constraint 
$\upsilon\geq\upsilon_{\min}(\eta)$, we obtain (\ref{optupseta}).

Thus, we now show that 
$f_\eta(\upsilon)$ is a decreasing function of $\upsilon$. We have
\begin{align}
&(1+\zeta(\upsilon,\eta))^2\frac{\mathrm df_\eta(\upsilon)}{\mathrm d\upsilon}
\nonumber\\&
=
- (\eta-1)\frac{(\eta-1)(\upsilon+\eta/2-1)+2\eta}{2\upsilon\eta^2}(1+\zeta(\upsilon,\eta))
\nonumber\\&
+[\upsilon-\hat u_{\mathrm{comm}}(\upsilon,\eta)]\frac{(\eta-1)(\eta/2-1)+2\eta}{2\upsilon^2\eta}(1+\zeta(\upsilon,\eta))
\nonumber\\&
+[\upsilon-\hat u_{\mathrm{comm}}(\upsilon,\eta)]\frac{(\eta-1)(\upsilon+\eta/2-1)+2\eta}{2\upsilon\eta}\zeta^\prime(\upsilon,\eta)
\nonumber\\&
- \frac{(\eta-1)}{\eta}\zeta(\upsilon,\eta)(1+\zeta(\upsilon,\eta))
- \frac{(\eta-1)(\upsilon+\eta/2-1)}{\eta}\zeta^\prime(\upsilon,\eta)
\nonumber\\&
+\eta\zeta^\prime(\upsilon,\eta)(1+\zeta(\upsilon,\eta))
\nonumber\\&
+[1/\upsilon-1/\hat u_{\mathrm{comm}}(\upsilon,\eta)/\eta](\eta/2+1)(1+\zeta(\upsilon,\eta))^2,
\end{align}
where $\zeta^\prime(\upsilon,\eta)=\frac{\mathrm d\zeta(\upsilon,\eta)}{\mathrm d\upsilon}$.
By simplifying and reorganizing the expression, we obtain
\begin{align}
% THIS ONE IS CORRECT!
% \eta[\upsilon-\hat u_{\mathrm{comm}}(\upsilon,\eta)]+[\eta^2/2+3/2\eta-1]=(\eta-1)\upsilon
%
% (\eta-1)(\upsilon+\eta/2-1)=\eta[\upsilon-\hat u_{\mathrm{comm}}(\upsilon,\eta)]+\eta^2
%
% (\eta-1)\hat u_{\mathrm{comm}}(\upsilon,\eta)
% =[\upsilon-\hat u_{\mathrm{comm}}(\upsilon,\eta)]+(\eta^2/2+3/2\eta-1)
%
&(1+\zeta(\upsilon,\eta))^2\frac{\mathrm df(\upsilon)}{\mathrm d\upsilon}
=
%% OK VERIFIED!!
-\frac{1}{\upsilon}[\upsilon-\hat u_{\mathrm{comm}}(\upsilon,\eta)]^2
\nonumber\\&
\times
\left[
\frac{\eta+2}{2\eta\upsilon\hat u_{\mathrm{comm}}(\upsilon,\eta)}
+\frac{\upsilon-\hat u_{\mathrm{comm}}(\upsilon,\eta)}{4\upsilon(\upsilon+\eta/2-1)}
\right]
\\&
%% OK VERIFIED!!
- \zeta(\upsilon,\eta)\frac{[\upsilon-\hat u_{\mathrm{comm}}(\upsilon,\eta)]}{\upsilon^2}
\nonumber\\&
\times
\left[
\frac{\eta^2/2+3/2\eta-1}{\upsilon-\hat u_{\mathrm{comm}}(\upsilon,\eta)}
+\frac{\eta^3}{(\upsilon+\eta/2-1)(\eta-1)}
+\frac{(\eta+2)\upsilon}{\eta\hat u_{\mathrm{comm}}(\upsilon,\eta)}
\right]
\nonumber\\&
- \zeta(\upsilon,\eta)\frac{[\upsilon-\hat u_{\mathrm{comm}}(\upsilon,\eta)]^2}{\upsilon^2}
\left[
1+\frac{\eta^2+3\eta-2}{2(\upsilon+\eta/2-1)(\eta-1)}
\right]
%%%% OK VERIFIED!!
\nonumber\\&
- \zeta(\upsilon,\eta)^2\frac{1}{\upsilon}
\Bigg[
[\upsilon/\hat u_{\mathrm{comm}}(\upsilon,\eta)-1](1/2+1/\eta)
\nonumber\\&
+\frac{\eta^2}{(\upsilon+\eta/2-1)(\eta-1)}\frac{\eta^2/2+3/2\eta-1}{\upsilon-\hat u_{\mathrm{comm}}(\upsilon,\eta)}
\Bigg]
\nonumber\\&
- \zeta(\upsilon,\eta)^2\frac{1}{\upsilon}
\left[
\frac{\upsilon-\hat u_{\mathrm{comm}}(\upsilon,\eta)+2\eta}{(\upsilon+\eta/2-1)(\eta-1)}\eta^2
+1+\upsilon-\hat u_{\mathrm{comm}}(\upsilon,\eta)
\right]
<0,
\end{align}
where inequality holds since $\zeta(\upsilon,\eta)\geq 0$
and $\upsilon>\hat u_{\mathrm{comm}}(\upsilon,\eta)$.
This proves that $f_\eta(\upsilon)$ is strictly decreasing in $\upsilon$.

Now, note that, in the limit $\upsilon\to\frac{1}{2}\frac{\eta^2+3\eta-2}{\eta-1}$,
we obtain $\upsilon\to\hat u_{\mathrm{comm}}(\upsilon,\eta)$ and
 $\zeta(\upsilon,\eta)\to\infty$, yielding (\ref{limups}).
On the other hand, when $\upsilon=\upsilon_{\max}(\eta)$, 
by letting $x\triangleq\hat P_{\max}(\eta-1)[1+\sqrt{1+2\eta/\hat P_{\max}}]>0$ we obtain
\begin{align}
&f_\eta(\upsilon_{\max}(\eta))
=
- (\eta-1)\frac{x[\eta+2+x ]}{\eta^2+3\eta-2+2\eta x}
\\&
+\ln\left(1+\frac{(\eta-1)x}{\eta^2/2+3/2\eta-1+x}\right)(\eta/2+1)
\triangleq g(x).
\end{align}
The derivative of the above expression with respect to $x$ satisfies
\begin{align}
\frac{\mathrm d g(x)}{\mathrm dx}
\propto
- [\eta^2/2+3/2\eta-1][2\eta+2x+x\eta]
- x^2\eta
<0,
\end{align}
which satisfies the inequality since $\eta\geq 2$,
hence $g(x)$ is maximized at $x=0$, yielding
\begin{align}
f_\eta(\upsilon_{\max}(\eta))
=g(x)<g(0)=0.
\end{align}
The Theorem is thus proved.
\end{proof}
}

\end{document}